\documentclass[10pt,letterpaper]{article}
\usepackage[top=0.85in,left=1.35in,footskip=0.75in,marginparwidth=2in]{geometry}
\usepackage{amsmath,amsfonts,amssymb}
\usepackage{graphicx}
\usepackage[colorlinks=true, allcolors=blue]{hyperref}
\usepackage{subcaption}
\usepackage[utf8]{inputenc}

\usepackage{cite}

\usepackage{nameref,hyperref}

\usepackage[right]{lineno}

\usepackage{microtype}
\DisableLigatures[f]{encoding = *, family = * }

\raggedright
\usepackage{changepage}

\usepackage[aboveskip=1pt,labelfont=bf,labelsep=period,singlelinecheck=off]{caption}

\makeatletter
\renewcommand{\@biblabel}[1]{\quad#1.}
\makeatother

\usepackage{lastpage,fancyhdr,graphicx}
\usepackage{epstopdf}
\fancyheadoffset[L]{2.25in}
\fancyfootoffset[L]{2.25in}

\usepackage{color}

\definecolor{Gray}{gray}{.25}

\usepackage{graphicx}

\usepackage{sidecap}

\usepackage{wrapfig}
\usepackage[pscoord]{eso-pic}
\usepackage[fulladjust]{marginnote}
\reversemarginpar

\begin{document}
\vspace*{0.35in}

\begin{flushleft}
{\Large
\textbf\newline{Identifying chromophore fingerprints of brain tumor tissue on hyperspectral imaging using principal component analysis}
}
\newline
\\
Ivan Ezhov\textsuperscript{1,*},
Luca Giannoni\textsuperscript{2,3},
Suprosanna Shit\textsuperscript{1},
Frederic Lange\textsuperscript{6},
Florian Kofler\textsuperscript{1,4},
Bjoern Menze\textsuperscript{5},
Ilias Tachtsidis\textsuperscript{6},
Daniel Rueckert\textsuperscript{1,7}
\\
\bigskip
\bf{1} Klinikum rechts der Isar, Technical University of Munich, Munich
\\
\bf{2} University of Florence, Florence
\\
\bf{3} European Laboratory for Non-Linear Spectroscopy, Florence
\\
\bf{4} Helmholtz AI, Helmholtz Zentrum München, Munich
\\
\bf{5} University of Zurich, Zurich
\\
\bf{6} University College London, London
\\
\bf{7} Imperial College London, London 
\\

\bigskip
* ivan.ezhov@tum.de

\end{flushleft}

\section*{Abstract}
Hyperspectral imaging (HSI) is an optical technique that processes the electromagnetic spectrum at a multitude of monochromatic, adjacent frequency bands. The wide-bandwidth spectral signature of a target object's reflectance allows fingerprinting its physical, biochemical, and physiological properties. HSI has been applied for various applications, such as remote sensing and biological tissue analysis. Recently, HSI was also used to differentiate between healthy and pathological tissue under operative conditions in a surgery room on patients diagnosed with brain tumors. In this article, we perform a statistical analysis of the brain tumor patients' HSI scans from the HELICoiD dataset with the aim of identifying the correlation between reflectance spectra and absorption spectra of tissue chromophores. By using the principal component analysis (PCA), we determine the most relevant spectral features for intra- and inter-tissue class differentiation. Furthermore, we demonstrate that such spectral features are correlated with the spectra of cytochrome, i.e., the chromophore highly involved in (hyper) metabolic processes. 
Identifying such fingerprints of chromophores in reflectance spectra is a key step for automated molecular profiling and, eventually, expert-free biomarker discovery.


\section*{Introduction}
Hyperspectral imaging is a noninvasive optical sensing technique that uses a broad range of narrow wavelength bands to analyze a target object \cite{khan2018modern}. A collection of reflections from all the bands can serve as a fingerprint of various physical, chemical, and physiological properties of matter. HSI has proven useful for remote sensing \cite{govender2007review}, drug screening \cite{puchert2010near}, and medical applications \cite{lu2014medical}. Recently the imaging technique has also been used to identify functional and pathological biomarkers of brain tissue \cite{giannoni2018hyperspectral,fabelo2019vivo}. 
Applied to biological tissues, HSI facilitates identifying biomarkers such as tissue metabolic activity or oxygenation. In turn, the biomarkers can shed light on the functional and pathological state of the examined tissue \cite{hyperprobe}. 

Differentiation between biomarkers from the reflection spectra would in large benefit by relating the spectra with absorption and scattering of the incoming light.
Absorption and scattering are the two main processes for light energy dissipation. Mathematically, one can describe their effect on the incoming electromagnetic wave via the Beer-Lambert law:

\begin{equation}
I_R(\lambda) = I_0(\lambda) e^{-(\mu_a(\lambda)+\mu_s(\lambda))\cdot l}
\end{equation}

where $I_0(\lambda)$ is the intensity of the incoming light, $I_R(\lambda)$ is the intensity of the reflected light captured by the detector camera, $\mu_a$ and $\mu_s$ are the absorption and scattering coefficients, $\lambda$ is the wavelength, and $l$ is the light pathlength.
The scattering process originates from different structural inhomogeneities in living tissue. 
The scattering coefficient $\mu_s$ is often analytically described using a low-degree polynomial dependency on the wavelength $\mu_s \sim \lambda^{(-n)}$, with $n$ being the degree \cite{jacques2013optical}. 
Different from the scattering, the absorption is not a bulk effect but rather occurs at the level of light interaction with single tissue molecules (or, more precisely, with molecules' chromophores). The energy dissipation due to absorption is transformed into the excitation of molecules. Since the excitation happens when light energy matches the distance between quantum energy states (which is a unique molecular property), chromophores' absorption spectra possess characteristic peaks.  

Brain tissues vary in their content of chromophores. For example, blood vessels have a relatively larger concentration of hemoglobin, whereas glioma tissue presumably has a higher percentage of cytochrome (a protein actively involved in metabolic processes) \cite{abramczyk2021revision,zong2016mitochondria}. Thus the total absorption spectra ($\sim exp(-(\sum_i c_i\mu_a^i$)) should manifest varying spectral signatures (here $c_i$ defines the concentration of a particular chromophore). Correspondingly, the captured reflection spectrum varies across tissue types as it is inversely proportional to the total absorption. The open question is whether one can solve the inverse problem, i.e., retrieve from the reflectance spectrum the composition of chromophores.

Several works exist attempting to perform unmixing of a reflection spectrum into a composition of chromophores spectra \cite{dobigeon2013nonlinear,dobigeon2016linear}. However, the main bottleneck of recovering a physiologically complete chromophore set is the ill-posedness of the inverse problem. Despite having characteristic peaks, the absorption spectra do not form an orthogonal basis within the HSI operation range of wavelengths. Thus, mathematically speaking, the mapping between reflection spectra and chromophores set is not bijective, i.e., different combinations of chromophores absorption can equally fit the reflectance. This is one of the reasons why existing works test unmixing algorithms with a limited number of chromophores in a composition. 

Inspired by previous works \cite{yokoyama2003estimation,gerstner2012hyperspectral}, in this article, we aim to identify chromophores spectra from glioma HSI images in a model-agnostic fashion by using statistical analysis means. Namely, we perform a PCA study to identify correlations between the principal components and the absorption spectra of various chromophores constituting brain tissues. 

\begin{figure}[t]
  \centering
  \includegraphics[width=0.98\linewidth]{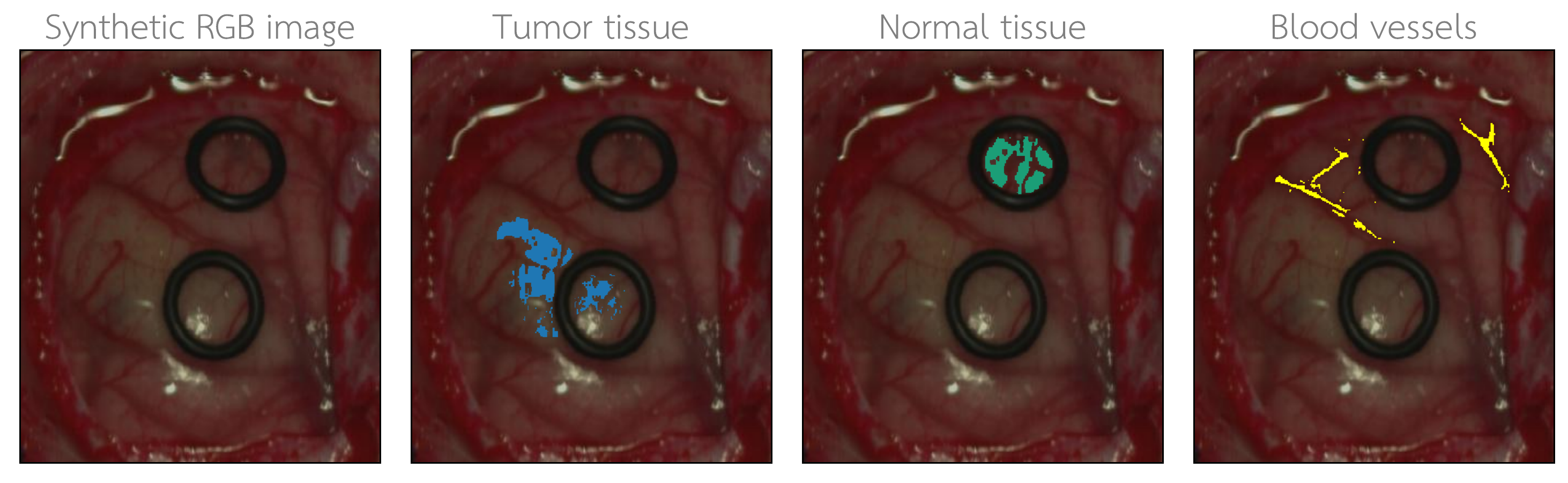}
  \caption{An example of a synthetic RGB image from the HELICoiD dataset. The image was obtained by merging three bands corresponding to red, green, and blue wavelengths from an HSI cube. The segmentations overlayed on top of the image represent three classes: tumor tissue, normal tissue, and blood vessels.}
\end{figure}

\begin{figure}[t]
\begin{subfigure}{.5\textwidth}
  \centering
  \includegraphics[width=0.94\linewidth]{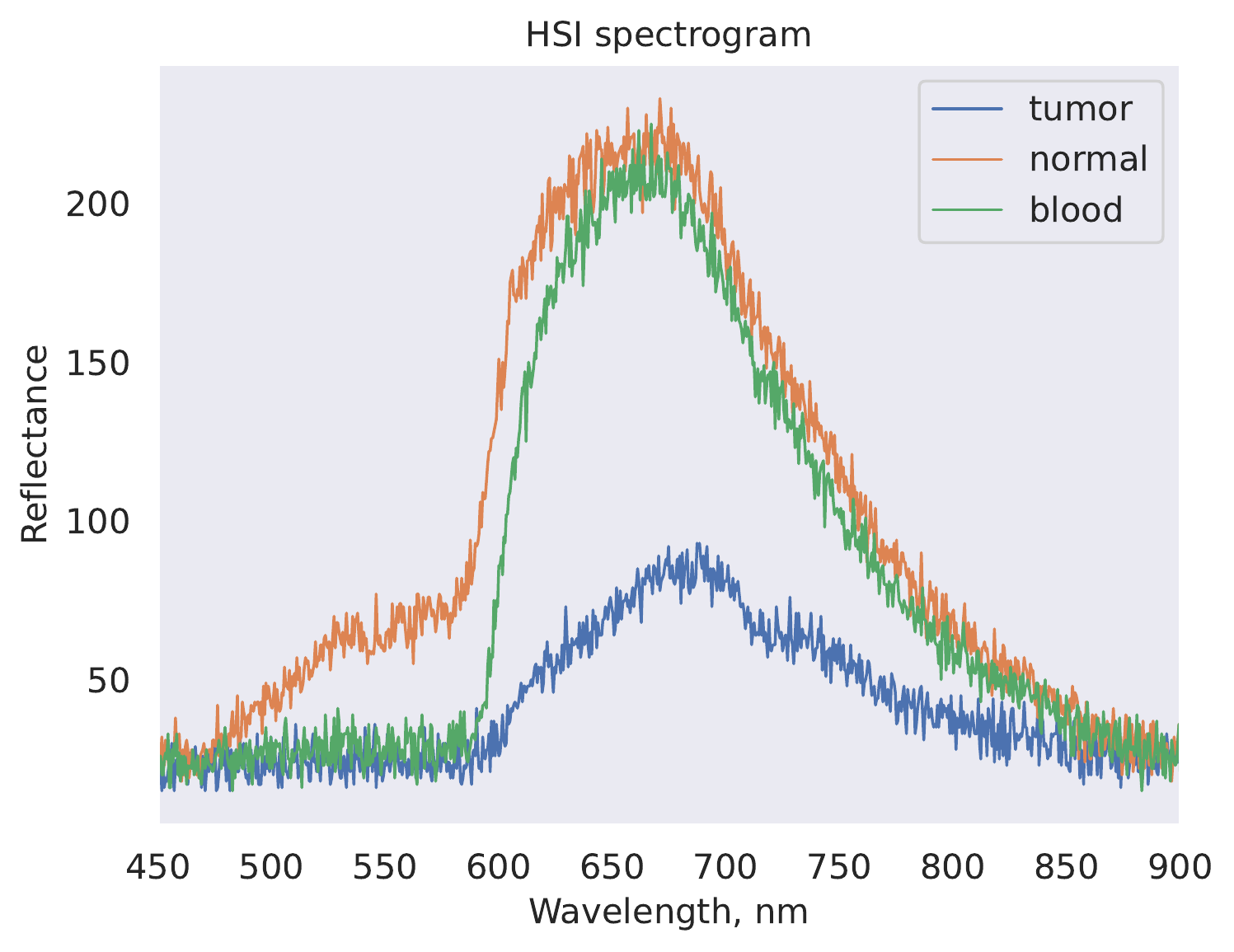}
  
  \label{fig:sfig1}
\end{subfigure}%
\begin{subfigure}{.5\textwidth}
  \centering
  \includegraphics[width=1.0\linewidth]{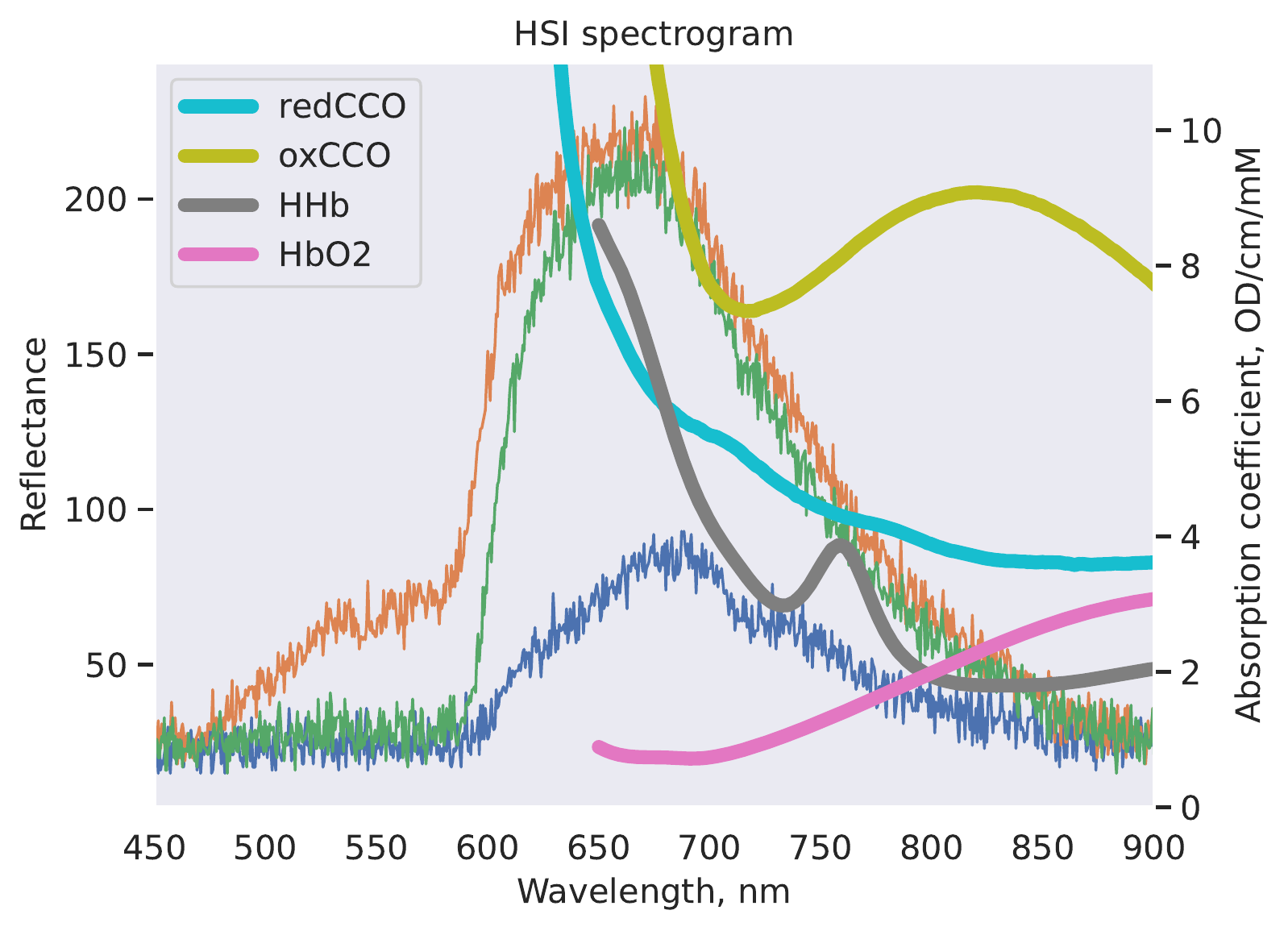}
  
  \label{fig:sfig2}
\end{subfigure}
\caption{Raw HSI spectra for three HELICoiD classes: tumor tissue, normal tissue, and blood vessels (left). Raw HSI spectra for the three HELICoiD classes and absorption spectra of typical chromophores: reduced and oxidised cytochrome-c-oxidase, oxy- and deoxy-hemoglobin (right).}
\end{figure}

\section{Method}

For our study, we used HSI images from the HELICoiD dataset \cite{fabelo2019vivo}.
The HELICoiD dataset consists of glioma patients which underwent HSI monitoring during surgical operations. The image dimensions are of varying spatial size across the dataset but with a fixed spectral size of 826 bands. The images were sparsely labeled (less than 25$\%$ of the image area) into four classes: tumor tissue, normal healthy tissue, blood vessels, and background, Figure 1. We preselected twelve patients which were diagnosed with grade IV glioblastoma as the primary tumor. From each of the preselected patients, we extracted spectra that belong to three classes (all HELICoiD classes, except the background). In total, we collected 30k spectra equally distributed over the three semantic classes. Figure 2 demonstrates typical raw spectral profiles for each class. The absorption spectra were taken from the BORL GitHub repository \cite{absorpgithub}.

Next, we performed the PCA for all 30k spectra in a high-dimensional space ($\mathbb{R}^{826}$) to identify axes of the highest variance. Our reasoning here is that, on one side, the projections of the first principal component (or a few first ones) into the original basis would inform us on how each HSI spectral band is important for capturing the data variance. On the other side, the spectral variance between the tissue classes originates from the different distribution of chromophores concentration. Therefore, we expect to observe a correlation between the principal components and the absorption spectra of chromophores.

\section{Results and discussion}

We performed PCA in two different settings:

1. First, we wanted to identify the principal components for a mixed dataset composed of spectra from different tissue classes. Such a test would allow determining spectral bands that best differentiate between the classes. Figures 3 and 4 show the results of such PCA tests. Here, \textit{ntb} denotes the 1st principal component for a dataset composed of all three classes, \textit{nt} - for normal and tumor tissue samples, \textit{nb} - for normal tissue and blood vessels. We visualize only the 1st component weights since it explains more than $98\%$ percent of the variance.

\begin{figure}[h]
\begin{subfigure}{.32\textwidth}
  \centering
  \includegraphics[width=1.0\linewidth]{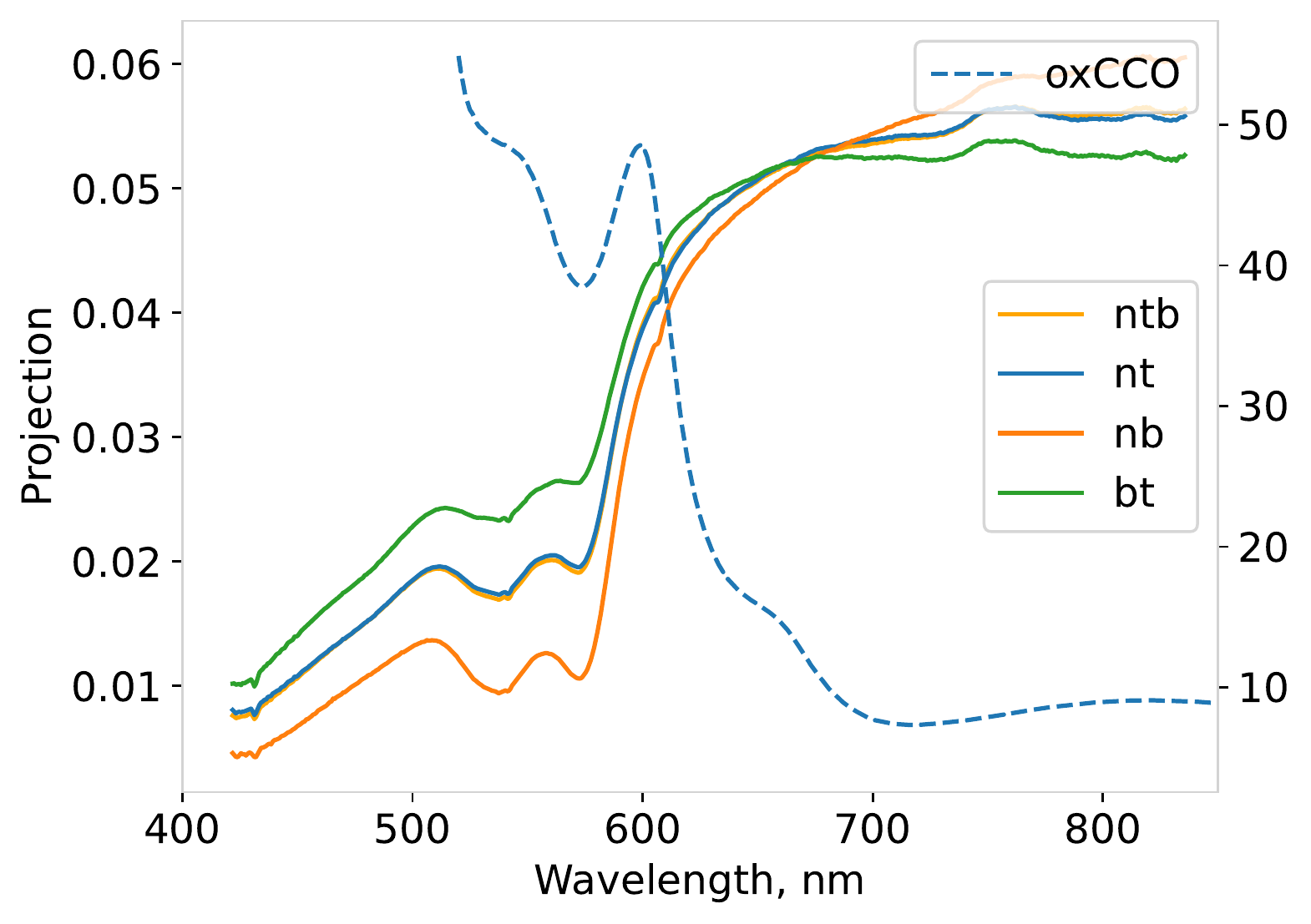}
  
  \label{fig:sfig1}
\end{subfigure}%
\begin{subfigure}{.31\textwidth}
  \centering
  \includegraphics[width=1.0\linewidth]{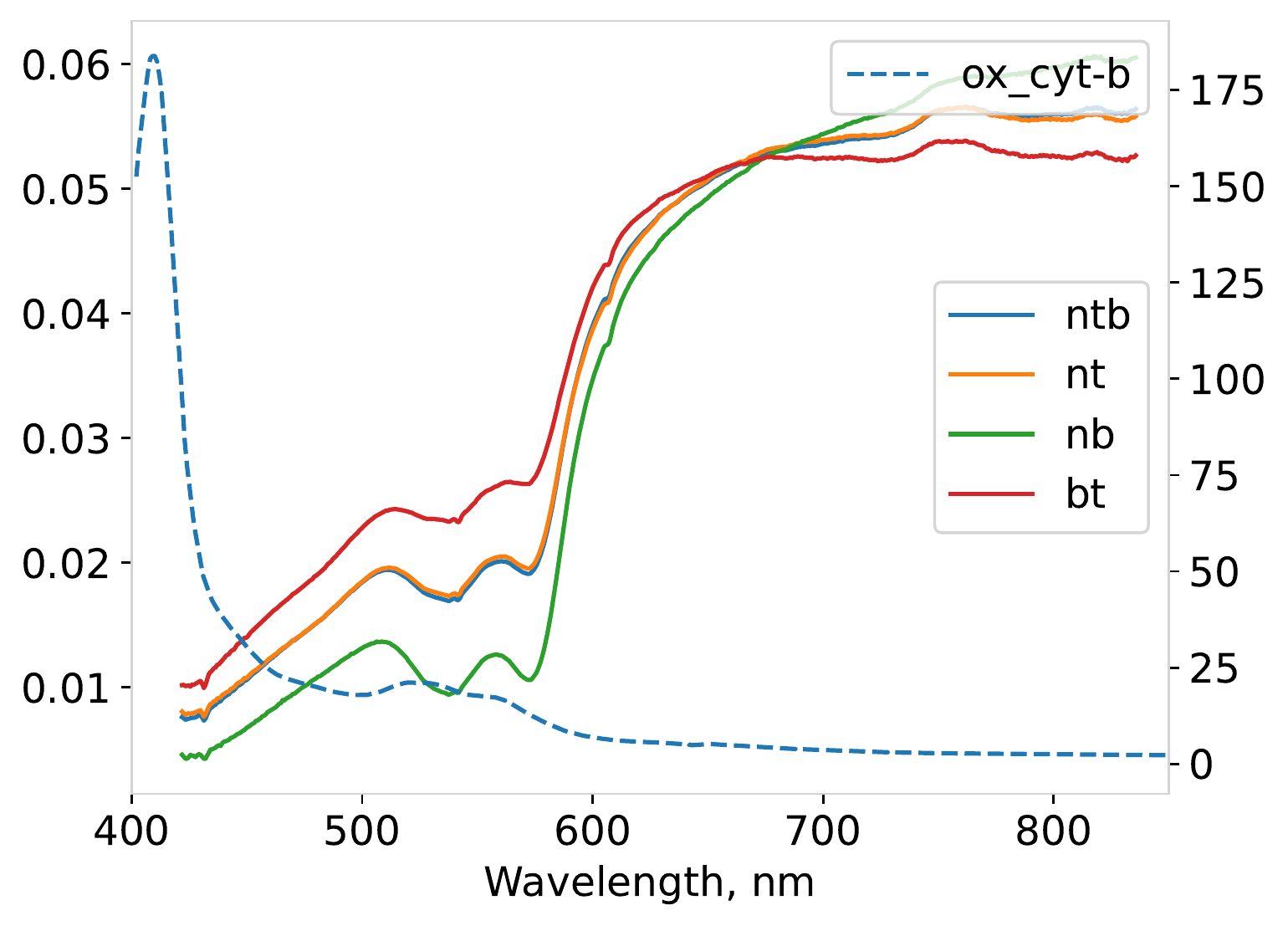}
  
  \label{fig:sfig2}
\end{subfigure}
\begin{subfigure}{.32\textwidth}
  \centering
  \includegraphics[width=1.0\linewidth]{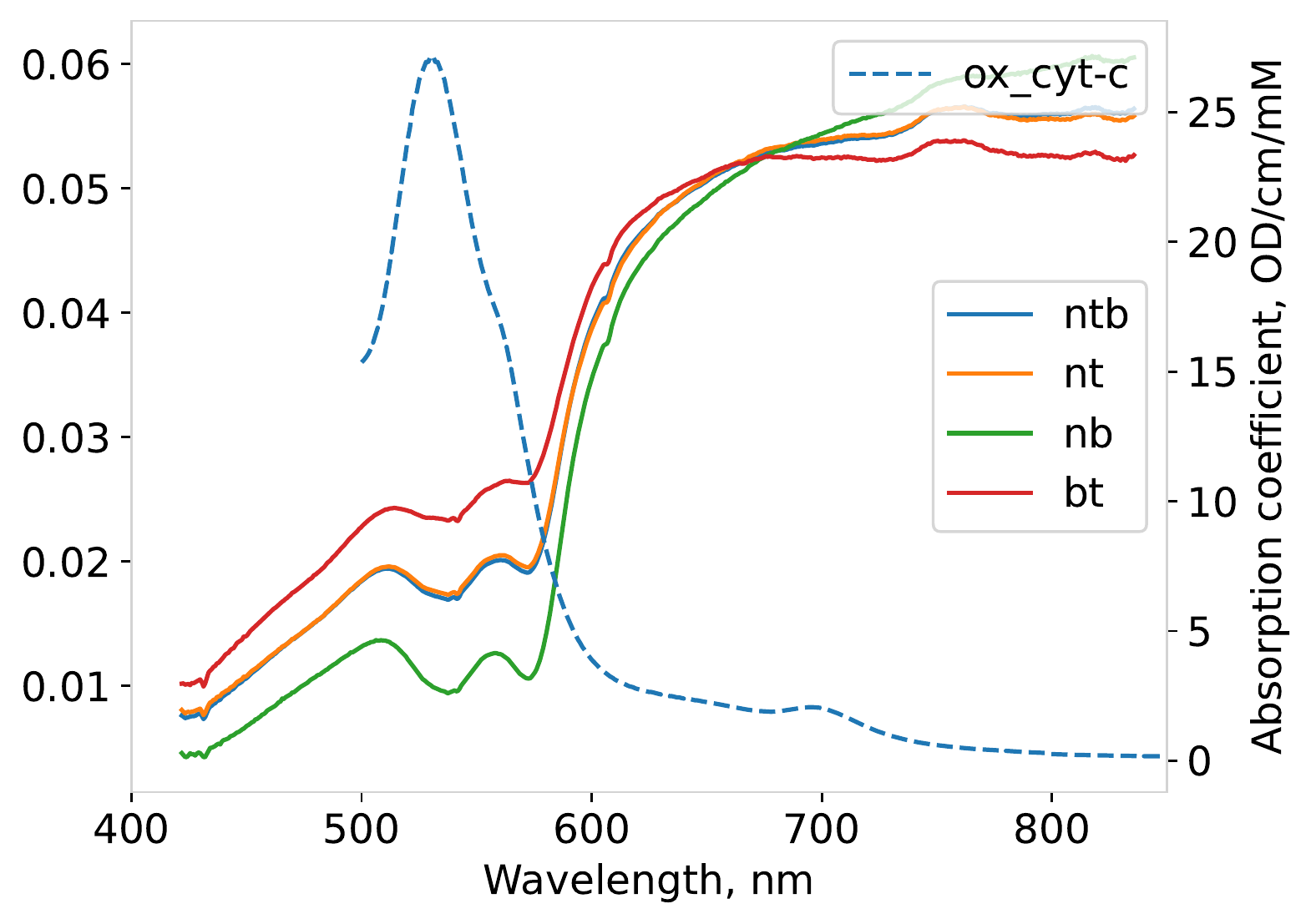}
  
  \label{fig:sfig2}
\end{subfigure}

\begin{subfigure}{.32\textwidth}
  \centering
  \includegraphics[width=1.0\linewidth]{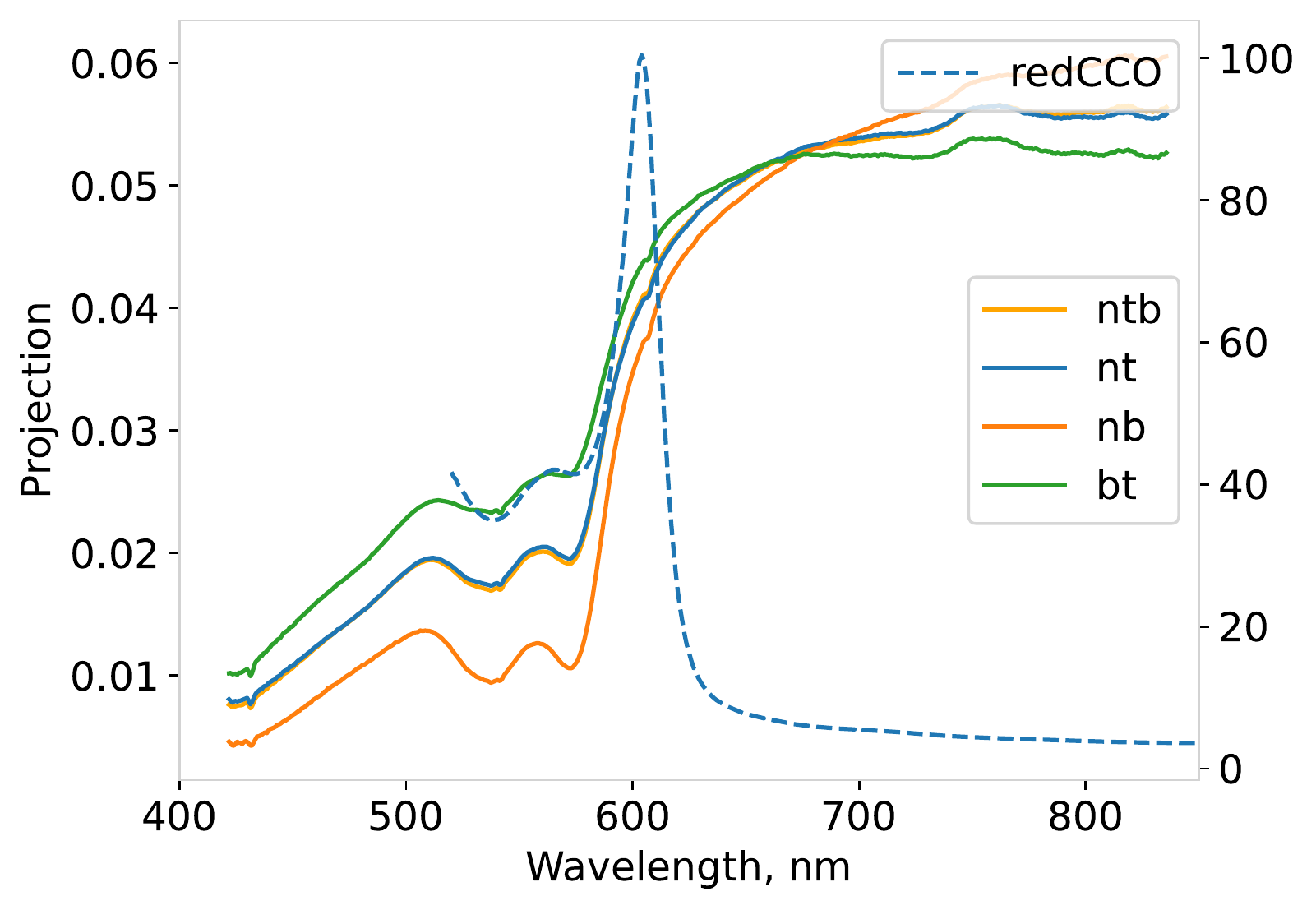}
  
  \label{fig:sfig1}
\end{subfigure}%
\begin{subfigure}{.31\textwidth}
  \centering
  \includegraphics[width=1.0\linewidth]{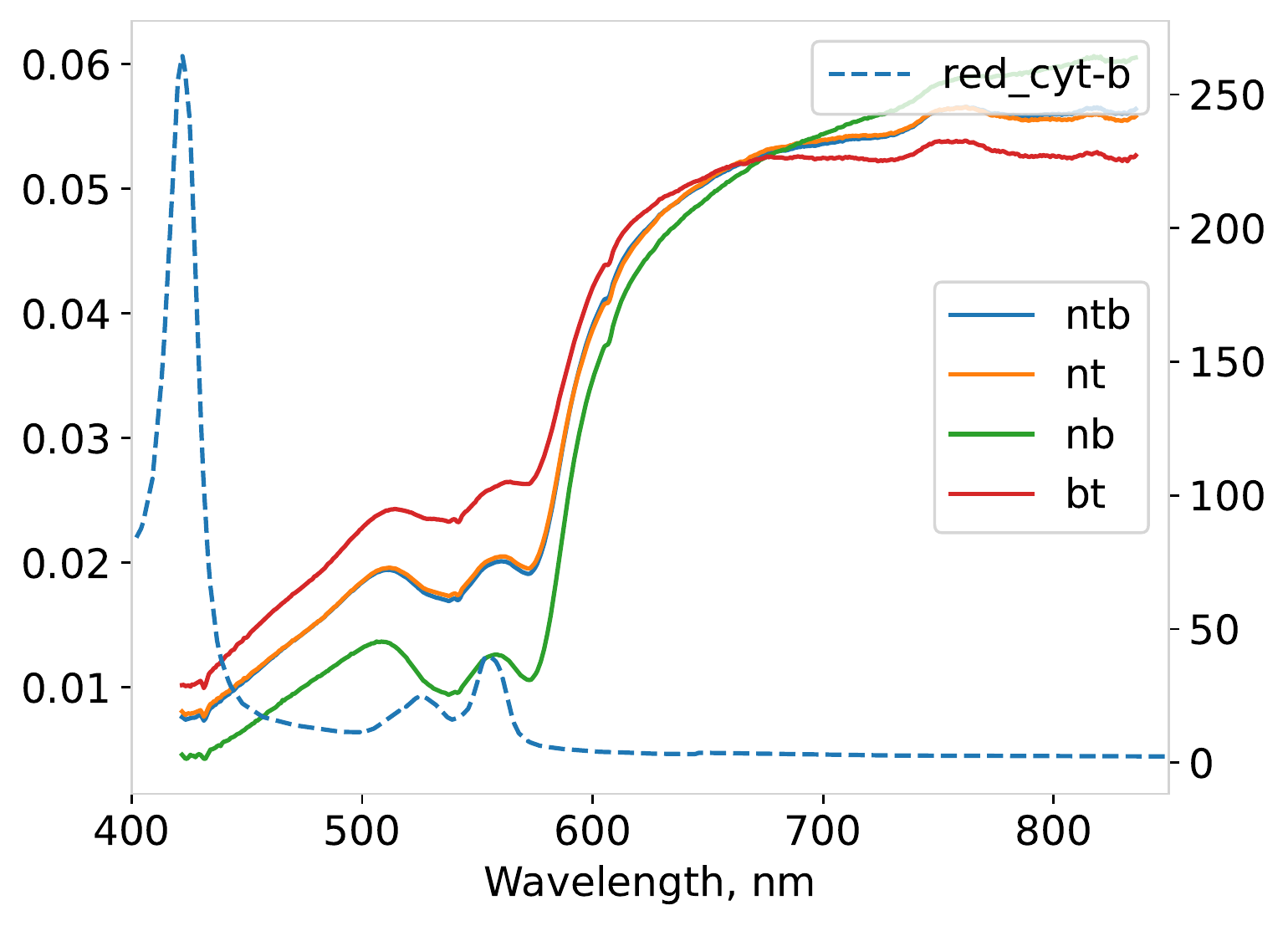}
  
  \label{fig:sfig2}
\end{subfigure}
\begin{subfigure}{.32\textwidth}
  \centering
  \includegraphics[width=1.0\linewidth]{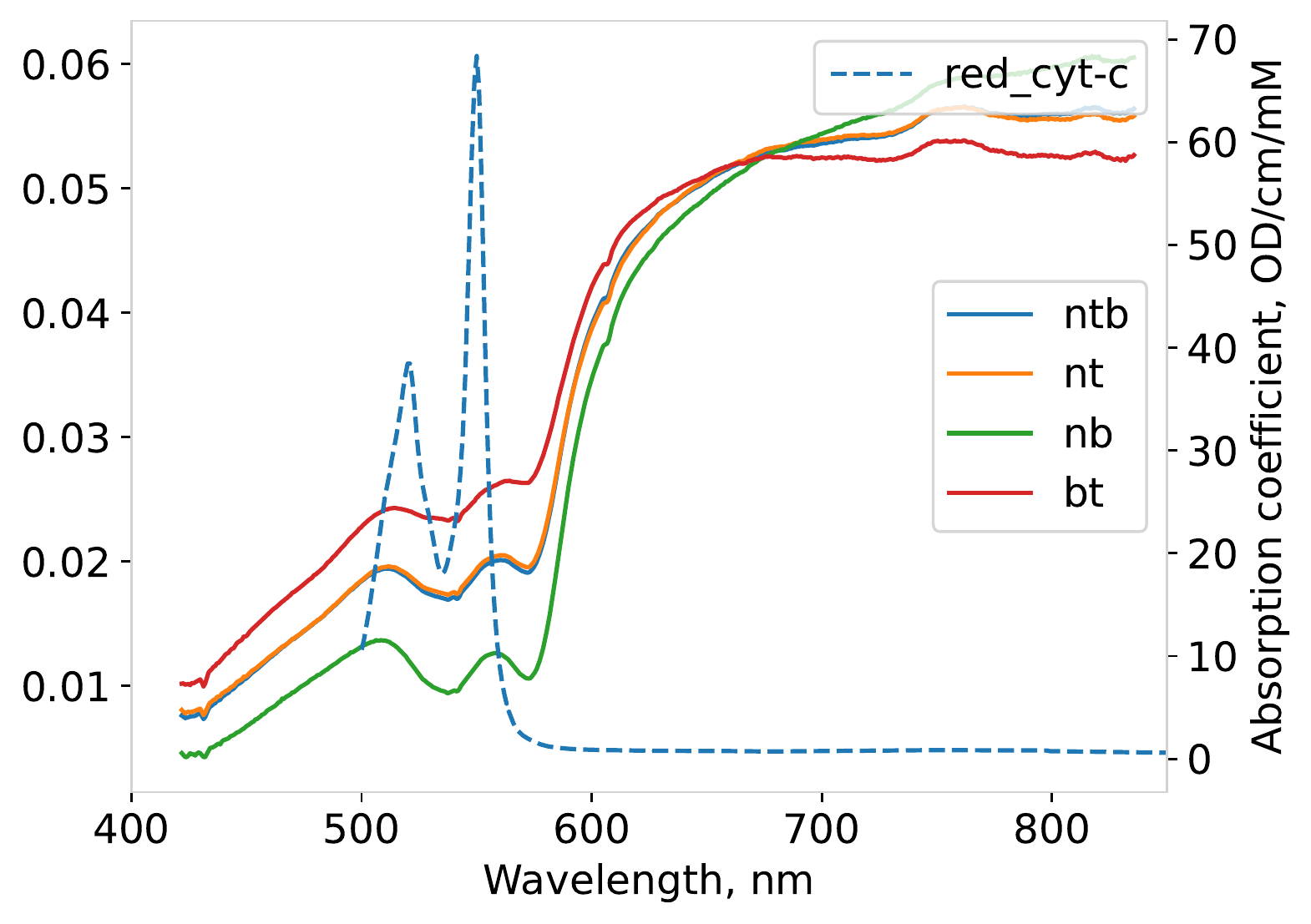}

  \label{fig:sfig2}
\end{subfigure}\\
\caption{Absorption spectra of cytochromes: oxidized (upper row) and reduced (bottom row) of three prosthetic group types (CCO, cyt-b, and cyt-c). In solid line, we show the 1st principal component for four different datasets: "ntb" denotes a dataset composed of all three classes (normal tissue, tumor, blood vessels), "nt" - is for normal and tumor tissue samples, "nb" - for normal tissue and blood vessels, and "bt" - for blood vessels and tumor. Reduced cytochrome-c-oxidase (redCCO) reveals the highest correlation with the principal component.}
\label{fig:fig}
\end{figure}

As evident from the figures, the range between 500 and 600 nm brings the highest correlation. Particularly absorption profile of the reduced cytochrome-c-oxidase (redCCO) reveals a very close match with the principal component. This can confirm our original hypothesis, as this is the interval of wavelengths where redCCO has characteristic absorption peaks. Cytochromes are present in a high concentration in the tumor microenvironment, less so in normal tissue, and only marginally in the endothelial cells of the inner walls of blood vessels. Our statistical analysis accurately captures this biological fact - the 1st principal component has the highest weights for separation between tumor tissue and blood vessels (\textit{bt}) and tumor against normal tissue (\textit{nt}) while having smaller weights for the blood versus normal tissue (\textit{nb}) separation.

\begin{figure}[h]
\centering
\begin{subfigure}{.32\textwidth}
  \centering
  \includegraphics[width=1.0\linewidth]{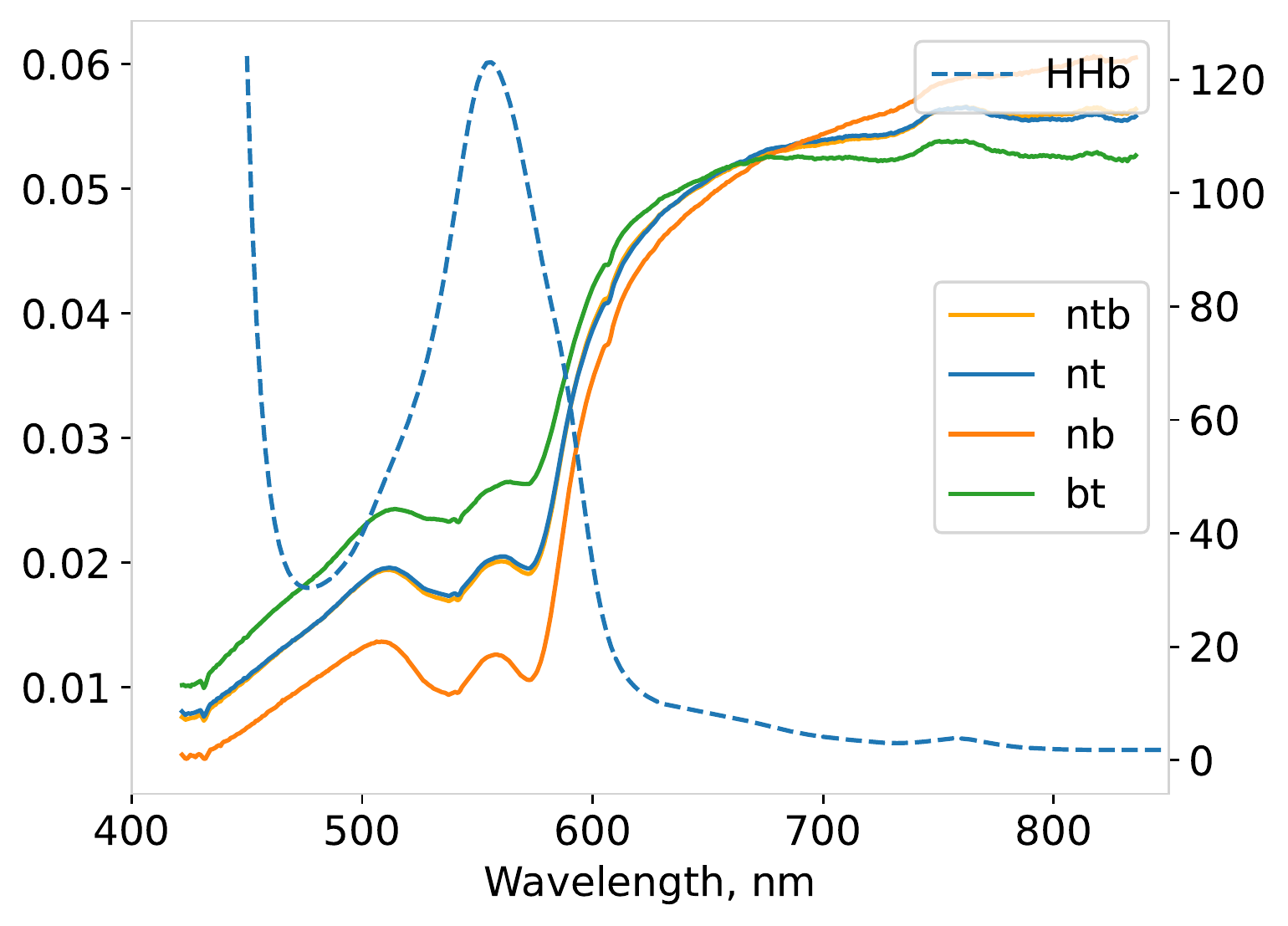}  
\end{subfigure}
\begin{subfigure}{.31\textwidth}
  \centering
  \includegraphics[width=1.0\linewidth]{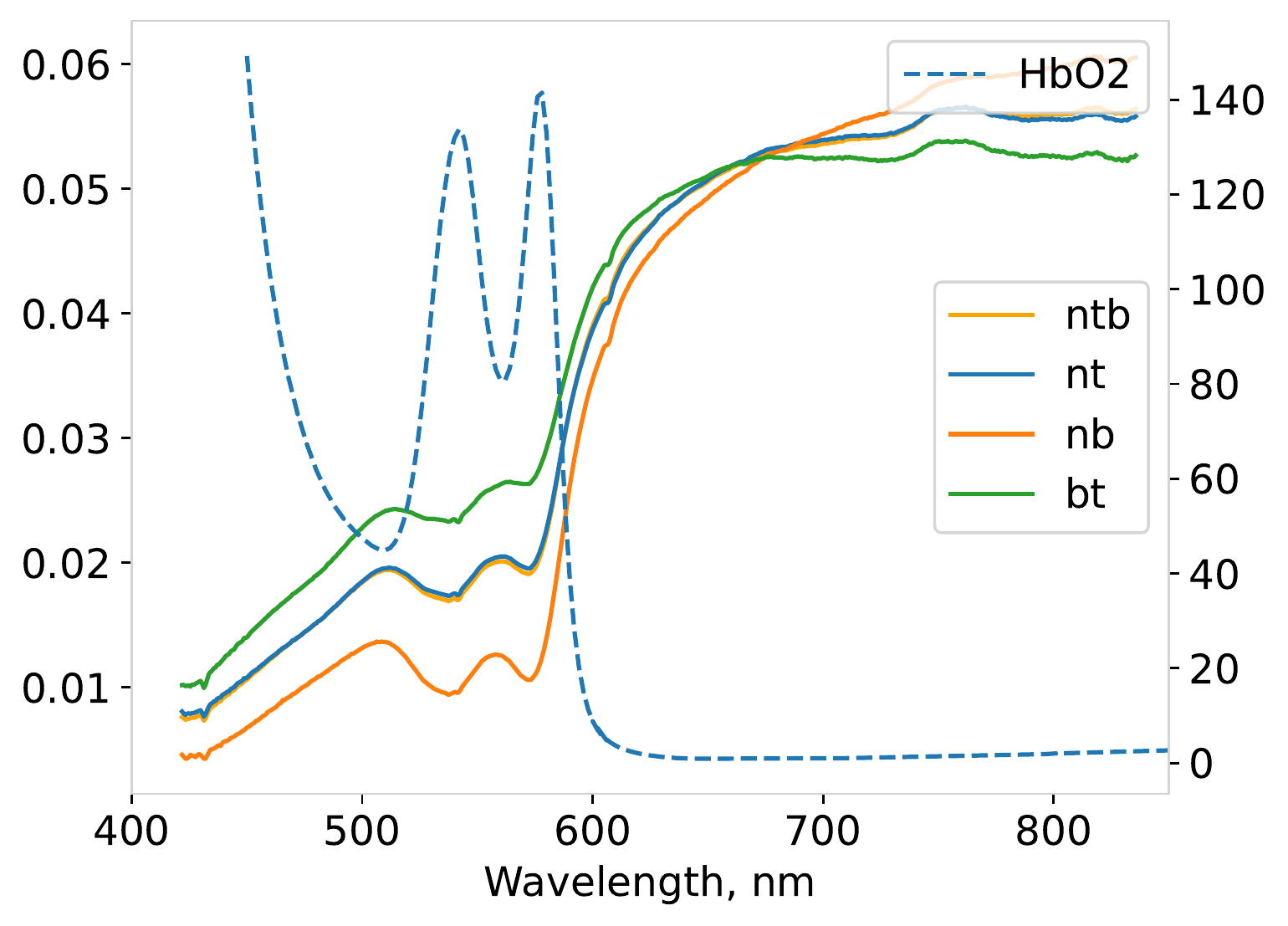}
\end{subfigure}
\begin{subfigure}{.32\textwidth}
  \centering
  \includegraphics[width=1.0\linewidth]{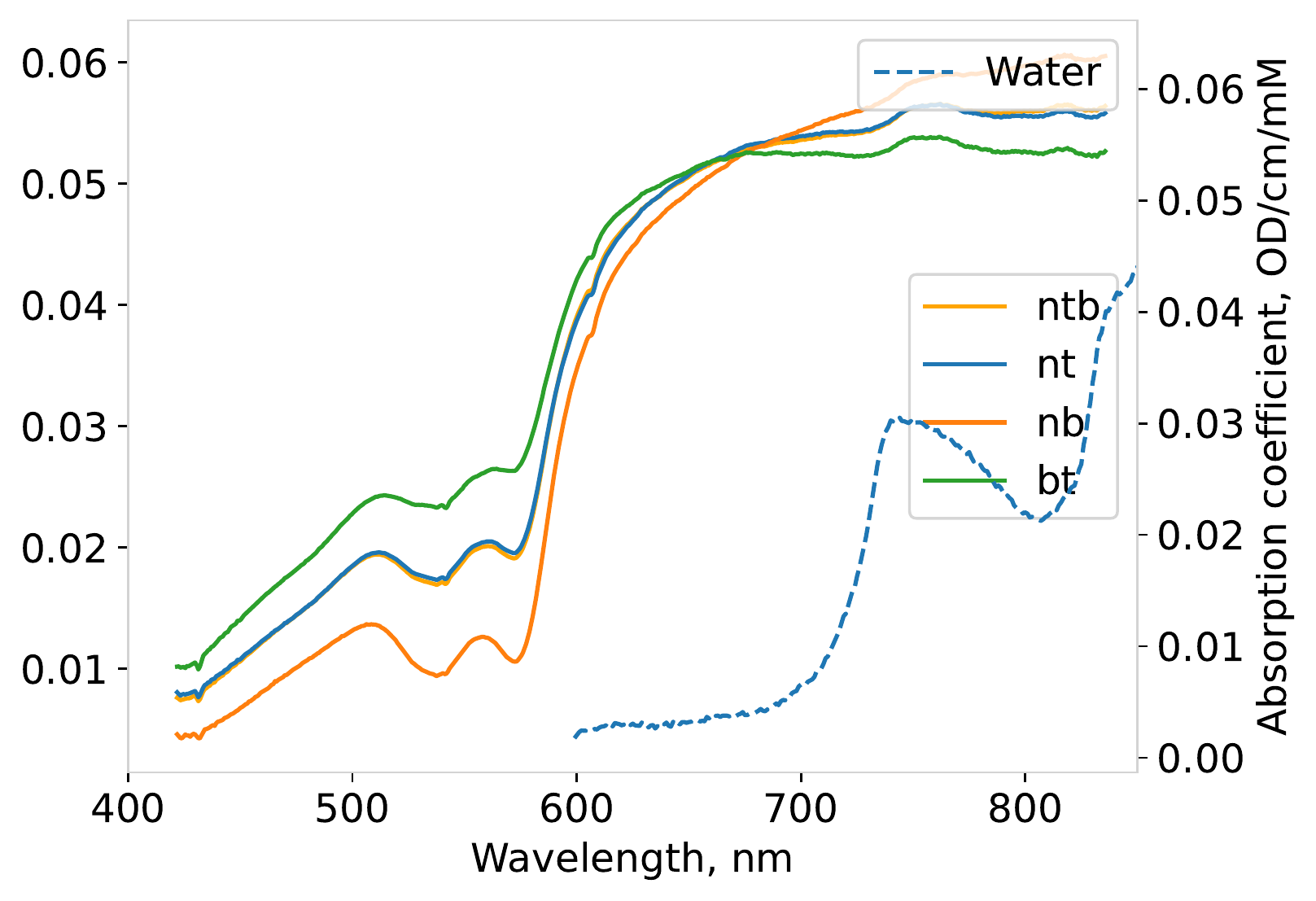}
\end{subfigure}\\
\caption{Absorption spectra of hemoglobin, deoxy- (left) and oxy- (middle), and water(right). In solid line, we show the 1st principal component for four different datasets: "ntb" denotes a dataset composed of all three classes (normal tissue, tumor, blood vessels), "nt" - is for normal and tumor tissue samples, "nb" - for normal tissue and blood vessels, and "bt" - for blood vessels and tumor.}

\end{figure}

\begin{figure}[h]
\centering
\includegraphics[width=0.5\linewidth]{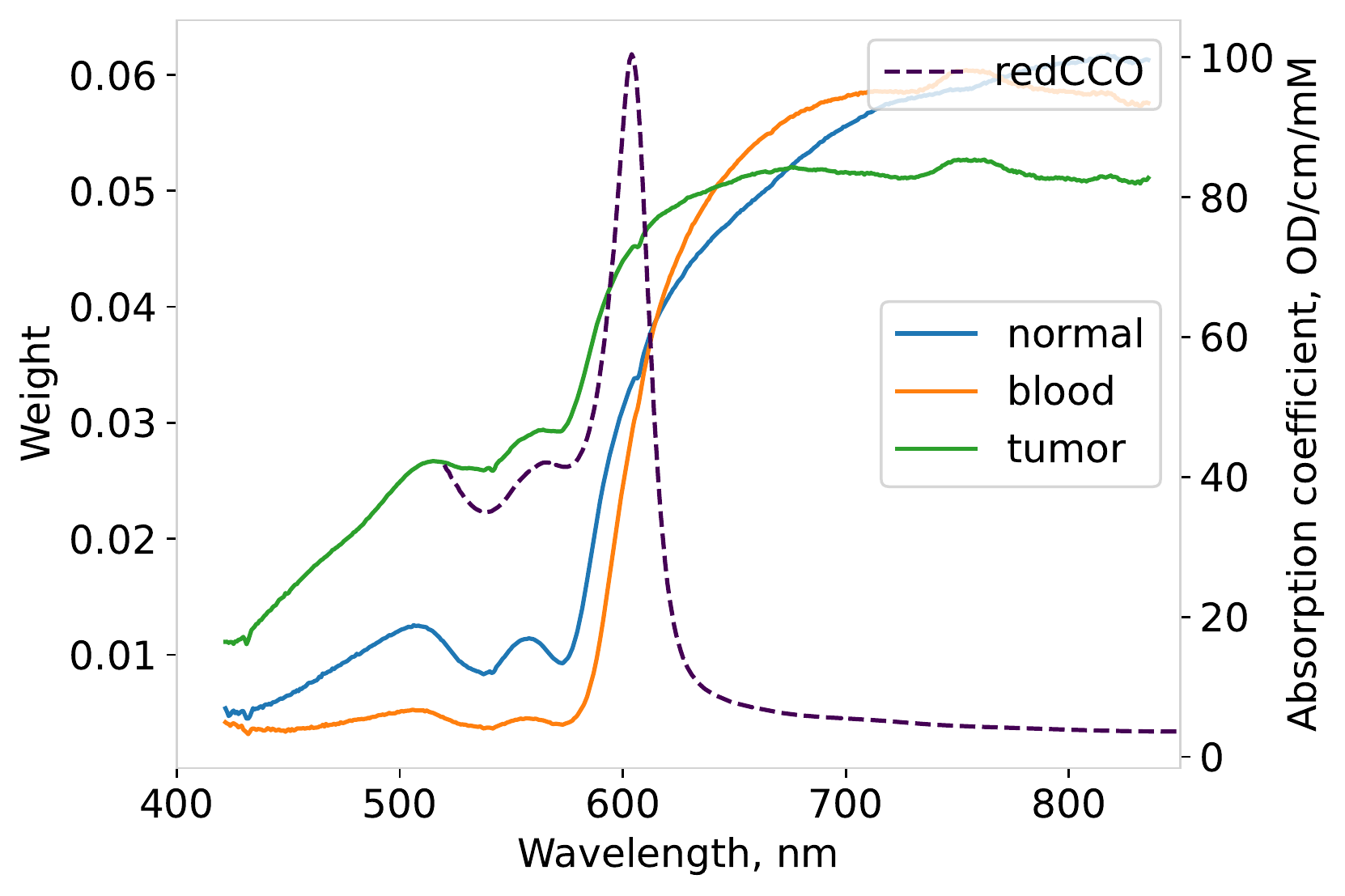}
\caption{Absorption spectra of reduced cytochrome and PCA components. In solid line, we show the 1st principal component for four different datasets: "ntb" denotes a dataset composed of all three classes (normal tissue, tumor, blood vessels), "nt" - is for normal and tumor tissue samples, "nb" - for normal tissue and blood vessels, and "bt" - for blood vessels and tumor.}
\end{figure}

2. We wanted to test whether PCA can reveal spectral signatures correlated with molecular absorption within a single class. This test is motivated by the fact that glioma tissue possesses high variability of cytochrome concentration since the tumor is vastly heterogeneous. In enhancing actively proliferating tumor, the hypermetabolism should be accompanied by an abnormal cytochrome amount \cite{abramczyk2021revision}. In contrast, no proliferation is expected in the necrotic core area, and thus, the concentration of cytochrome should be minimal. Therefore the weights of the 1st principal component are expected to be aligned with the cytochrome absorption and be more pronounced for the tumor class than for healthy tissue and vessels. This was also confirmed by the PCA, as seen in Figure 5 - the 1st principal component of the dataset composed of tumor samples has the highest weight in the 500-600 nm range. 
We want to point out that total absorption from any tissue class is a combination of absorption from a set of chromophores. For example, Figure 4 illustrates that oxy- and deoxyhemoglobin also have characteristic peaks in this interval. Hemoglobin concentration in tissues, though, has a contrary distribution to cytochrome, being higher in blood vessels and less in tumor and normal tissues. However, as discussed just above, the intra-class PCA rather captures the relation between tissues in its dependency on cytochrome. This poses the question of whether the observation is due to a much higher concentration of cytochrome than hemoglobin in tissues or a much larger variance in cytochrome within a tissue class. We did not find evidence for the former in literature, rather opposite is typically observed \cite{giannoni2020investigation}.
And if the latter is true, the difference between the magnitude of the principal components for the three tissue classes can quantify the intra-class chromophore variance. Such knowledge helps in understanding plausible ranges of concentration which is in turn valuable for deciphering the molecular profiling of tissue.

\section{Conclusion}
In this work, we analyze HSI glioma images from the HELICoiD project. We perform a PCA-based statistical analysis of this dataset to identify chromophore absorption signatures in the HSI reflection spectra. PCA revealed the correlation of chromophores, especially cytochrome, with the principal components. We discuss the possibility of using such analysis to decypher relative chromophore concentration in brain tissues. 
We see such analysis as a vital tool for the identification of chromophore fingerprints complementing traditional spectral unmixing techniques. 


\section{Acknowledgments} 
 
The authors have received funding from the European Union’s Horizon Europe research and innovation program under grant agreement number 101071040. F.L. and I.T. are supported by UCL, which, as a UK participant in the EU HyperProbe project, is supported by UKRI grant number 10048387. B.M. is supported by Helmut Horten Foundation.

\bibliography{library}

\bibliographystyle{abbrv}

\end{document}